\documentclass[showpacs,10pt,twocolumn]{revtex4}
\usepackage[dvips]{graphicx}
\usepackage{amsmath}
\begin{document}
\title{Secure quantum dense coding via tripartite entangled GHZ
state in cavity QED}

\author{Zheng-Yuan Xue}
\email{zyxue@ahu.edu.cn}
\author{You-Min Yi}
\author{Zhuo-Liang Cao}
\email{zlcao@ahu.edu.cn}

\affiliation{Anhui Key Laboratory of Information Material {\&}
Devices, School of Physics {\&} Material Science, Anhui University,
Hefei, 230039, P R China}
\begin{abstract}
We investigate economic protocol to securely encoding classical
information among three users via entangled GHZ states. We implement
the scheme in cavity QED with atomic qubits where the atoms interact
simultaneously with a highly detuned cavity mode with the assistance
of a classical field. The scheme is insensitive to the cavity decay
and the thermal field, thus based on cavity QED techniques presently
it might be realizable.
\end{abstract}
\pacs{03.67.Hk, 03.65.Ud, 42.50.Dv}

\maketitle

\section{Introduction}
Quantum entanglement, a fundamental feature of many-body quantum
mechanical systems, was seen as a key resource for many tasks in
quantum information processing. Among its novel applications,
quantum dense coding (QDC) \cite{1} and quantum secret sharing (QSS)
\cite{2} attract more and more public attention due to their
promising application in secure direct quantum communication. QDC is
a process to send two cbits of information, from a sender (Alice) to
a remote receiver (Bob), by sending only a single qubit. It works in
the following way. Initially, Alice and Bob shared a maximally
entangled state. The first step is an encoding process where Alice
performs one of the four local operations on her qubit, and then
sends the qubit to Bob. The second step is a decoding process. After
Bob received the qubit, he can discriminate the operation of Alice
using only local operations (Bell state measurement in the
pioneering work \cite{1}). Recently, much attention has been paid to
the studies of QDC both theoretically and experimentally.

In the realm of atom, cavity quantum electrodynamics (QED)
techniques has been proven to be a promising candidate for the
physical realization of quantum information processing. Recently,
many schemes have been proposed for quantum entanglement engineering
and quantum information processing \cite{3}. The cavity usually act
as memories in quantum information processing, thus the decoherence
of the cavity field becomes one of the main obstacles for the
implementation of quantum information in cavity QED. Recently, Zheng
and Guo proposed a novel scheme \cite{4}, which greatly prolong the
efficient decoherence time of the cavity. Osnaghi \textit{et al}.
\cite{5} had experimentally implemented the scheme using two Rydberg
atoms crossing a nonresonant cavity. Following the progress, schemes
for implementing QDC in cavity QED are also proposed \cite{lin,ye}.

QDC, despite for its novel classical capacity of sending classical
information, should be very deliberately used for the sake of
security of the process. If Bob is dishonest in the dense coding
process, then he can always successfully cheat Alice and use the
information willingly. Now, the question arises, is there any secure
QDC scheme exist? Fortunately, We note that QSS is likely to play a
important role in protecting secret quantum information. QSS is a
process securely distribute private key among many parties. With the
key, the sender, Alice can divide the message into two or more
shares. If and only if when they cooperate, they can get complete
information about the message. Meanwhile, if one of them is
dishonest, the honest players may keep the dishonest one from doing
any damage. Hence, after the pioneering work proposed by Hillery et
al. \cite{2}, QSS attracts a great deal of attentions in both
theoretical and experimental aspects.

In this paper, we investigate a secure QDC scheme with a tripartite
GHZ state via QSS in cavity QED. The sender (Alice) can transmit two
cbits of information by sending one qubit to one of the two
receivers, \textit{i.e}., Bob. By collaboration, they could obtain
the exact information of Alice, furthermore, any attempt to obtain
the secret information without cooperation cannot be succeed in a
deterministic way. Our scheme could work if one of them, and only
one, is not entirely trustful.

\section{QDC in cavity QED}
Suppose Alice wants to send secret information to a distant agent
Bob, she possesses three qubits and they are in the GHZ state. As
she does not know whether he is honest or not, she makes the
information shared by two users (\textit{i.e}., Bob and Charlie). If
and only if they collaborate, one of the users can read the
information, furthermore, individual users could not do any damage
to the process. Here, we base on the assumption that communication
over a classical channel is insecure, which means we cannot result
to the simplest method of teleportation \cite{25} to distribute the
information. Of course, one could also securely conquer the task
using standard quantum cryptography, but, on average, it requires
more resource and measurements \cite{26}. Alice can generate a
tripartite entangled GHZ state \cite{27} in cavity QED
\begin{equation}
\label{ghz}
|\psi\rangle_{1,2,3}=\frac{1}{\sqrt{2}}(|eee\rangle+i|gggg\rangle)_{1,2,3},
\end{equation}
where  $|e\rangle$ and $|g\rangle$  are the excited and ground
states of the atoms, respectively. Now, we present our scheme step
by step.

Step 1. Alice decides to select one of the following two possible
choices. With probability $p$, which is relatively very small, Alice
selects the first choice of security checking, which aims to check
the security of quantum channel, and the procedure continues to Step
2. Otherwise, Alice can decide to the information encoding step with
probability $(1-p)$, the aim of which is to encode  and implement
the QDC procedure. In this case, the procedure goes to Step 3.

Step 2. Security checking. Hillery \textit{et al}. \cite{2} show
that tripartite entangled GHZ state is sufficient to detect a
potential eavesdropper in the channel.

Step 3. Information encoding. After Alice confirms that Bob and
Charlie both receive their qubit from her, she performs one of the
four local operations $(\textit{I}, \sigma^{x},i\sigma^{y},
\sigma^{z})$  on her atom. These operations denote 2 cbits of
information, and will transform state in Eq. (\ref{ghz}) into
\begin{subequations}
\label{ghzu}
\begin{equation}
|\psi_{1}\rangle=\frac{1}{\sqrt{2}}(|eee\rangle+i|ggg\rangle)_{1,2,3},
\end{equation}
\begin{equation}
|\psi_{2}\rangle=\frac{1}{\sqrt{2}}(|gee\rangle+i|egg\rangle)_{1,2,3},
\end{equation}
\begin{equation}
|\psi_{3}\rangle=\frac{1}{\sqrt{2}}(|gee\rangle-i|egg\rangle)_{1,2,3},
\end{equation}
\begin{equation}
|\psi_{4}\rangle=\frac{1}{\sqrt{2}}(|eee\rangle-i|ggg\rangle)_{1,2,3}.
\end{equation}
\end{subequations}
Now the information is encoded into the pure
entangled state, which is shared among the three parties (Alice, Bob
and Charlie), the encoding of the two cbits information is
completed. Then Alice sends her atom to one of the two receivers
(\textit{i.e}., Bob), we will latter prove that which is the party
to send the atom to is not arbitrary. After a party receives the
atom, he/she will get hold of two of all the three atoms, and thus
he/she will have a higher probability of successful cheat compared
with the one who have not in the QDC process. So, Alice would send
her atom to the party, which is less likely to cheat.

Step 4. Information extracting. Assume Bob was selected to receive
Alice's atom, and then he has two atoms. Next, We consider two
identical two-level atoms simultaneously interacting with a
single-mode cavity and simultaneously driven by a classical field.
Then the interaction between the single-mode cavity and the two
driven atoms can be described, in the rotating-wave approximation,
as\cite{28}
\begin{eqnarray}
\label{abc} H=\omega_{0} S_{z} +\omega_{a} a^{+}a + \sum_{j=1}^{2}
\{g(a^{+}S_{j}^{-}+aS_{j}^{+})\nonumber\\
 +\Omega[ (S_{j}^{+}\exp
(-i\omega t)+S_{j}^{-}\exp (i\omega t)]\}
\end{eqnarray}
where $S_{z}=1/2\sum_{j=1}^{2} (|e\rangle_{j,j}\langle
e|-|g\rangle_{j,j}\langle g|)$, $S_{j}^{+}=|g\rangle_{j,j}\langle
e|$, $S_{j}^{-}=|e\rangle_{j,j}\langle g|$ and $|e\rangle_{j}$,
$|g\rangle_{j}$ are the excited and ground states of  $j$th atom,
respectively. $a^{+}$ and $a$ are the creation and annihilation
operators for the cavity mode, respectively. $g$ is the coupling
constant between cavity and particle, $\Omega$ is the Rabi
frequency, $\omega_{0}$ , $\omega_{a}$ and $\omega$ are atomic
transition frequency, cavity frequency and the frequency of the
driven classical field, respectively.

While the case of  $\omega_{0}=\omega$, the evolution operator of
the system, in the interaction picture, is
$U(t)=\exp(-iH_{0}t)\exp(-iH_{e}t)$, where
$H_{0}=\sum_{j=1}^{2}\Omega (S_{j}^{+}+S_{j}^{-})$, $H_{e}$ is the
effective Hamiltonian. In the strong driving regime $\Omega
\gg\delta, g$ ($\delta$ been the detuning between atomic transition
frequency $\omega_{0}$  and cavity frequency $\omega_{a}$) and the
case of $\delta \gg g$, there is no energy exchange between the
atomic system and the cavity thus the scheme is insensitive to both
the cavity decay and the thermal field. Then in the interaction
picture, the effective interaction Hamiltonian reads\cite{28}
\begin{eqnarray}
H_{e}=\frac{\lambda}{2}[\sum_{j=1}^{2}(|e\rangle_{j,j}\langle
e|+|g\rangle_{j,j}\langle g|)\nonumber\\
+\sum_{j=1;i\neq j}^{2}(S_{j}^{+}S_{k}^{+}+S_{j}^{+}S_{k}^{-}+H.c.)]
\end{eqnarray}
where $\lambda=g^{2}/2\delta$. If two atoms are simultaneously sent
into the cavity and simultaneously interact with it. Using the above
cavity, it is easy to verify the following evolvement
\begin{widetext}
\begin{subequations}
\label{te}
\begin{eqnarray}
|g\rangle|g\rangle\rightarrow e^{-i\lambda t}[\cos\lambda
t(\cos\Omega t|g\rangle-i\sin\Omega t|e\rangle)
(\cos\Omega t|g\rangle-i\sin\Omega t|e\rangle)\nonumber\\
-i\sin\lambda t(\cos\Omega t|e\rangle-i\sin\Omega t|g\rangle)
(\cos\Omega t|e\rangle-i\sin\Omega t|g\rangle)].
\end{eqnarray}
\begin{eqnarray}
|g\rangle|e\rangle\rightarrow e^{-i\lambda t}[\cos\lambda
t(\cos\Omega t|g\rangle-i\sin\Omega t|e\rangle)
(\cos\Omega t|e\rangle-i\sin\Omega t|g\rangle)\nonumber\\
-i\sin\lambda t(\cos\Omega t|e\rangle-i\sin\Omega t|g\rangle)
(\cos\Omega t|g\rangle-i\sin\Omega t|e\rangle)].
\end{eqnarray}
\begin{eqnarray}
|e\rangle|g\rangle\rightarrow e^{-i\lambda t}[\cos\lambda
t(\cos\Omega t|e\rangle-i\sin\Omega t|g\rangle)
(\cos\Omega t|g\rangle-i\sin\Omega t|e\rangle)\nonumber\\
-i\sin\lambda t(\cos\Omega t|g\rangle-i\sin\Omega t|e\rangle)
(\cos\Omega t|e\rangle-i\sin\Omega t|g\rangle)].
\end{eqnarray}
\begin{eqnarray}
|e\rangle|e\rangle\rightarrow e^{-i\lambda t}[\cos\lambda
t(\cos\Omega t|e\rangle-i\sin\Omega t|g\rangle)
(\cos\Omega t|e\rangle-i\sin\Omega t|g\rangle)\nonumber\\
-i\sin\lambda t(\cos\Omega t|g\rangle-i\sin\Omega t|e\rangle)
(\cos\Omega t|g\rangle-i\sin\Omega t|e\rangle)].
\end{eqnarray}
\end{subequations}
\end{widetext}
Choosing to adjust the interaction time  $\lambda
t=\pi/4$ and modulate the driving field $\Omega t=\pi$, lead the
quantum state of the three atoms system in Eq. (\ref{ghzu}), after
interaction, to
\begin{subequations}
\begin{eqnarray}
|\psi_{1}\rangle=\frac{1}{2}[|ee\rangle_{1,2}(|e\rangle+|g\rangle)_{3}
 -i|gg\rangle_{1,2}(|e\rangle-|g\rangle)_{3}],
\end{eqnarray}
\begin{eqnarray}
|\psi_{2}\rangle=\frac{1}{2}[|ge\rangle_{1,2}(|e\rangle+|g\rangle)_{3}
 -i|eg\rangle_{1,2}(|e\rangle-|g\rangle)_{3}],
 \end{eqnarray}
 \begin{eqnarray}
|\psi_{3}\rangle=\frac{1}{2}[-i|eg\rangle_{1,2}(|e\rangle+|g\rangle)_{3}
 +|ge\rangle_{1,2}(|e\rangle-|g\rangle)_{3}],
\end{eqnarray}
\begin{eqnarray}
|\psi_{4}\rangle=\frac{1}{2}[-i|gg\rangle_{1,2}(|e\rangle+|g\rangle)_{3}
+|ee\rangle_{1,2}(|e\rangle-|g\rangle)_{3}].
\end{eqnarray}
\end{subequations}
Obviously, one can see that there is an explicit correspondence
between Alice's operation and the measurements results of the two
receivers, which means that if they cooperate, both of them can get
the information. But if they do not choose to cooperate, neither of
the two users could obtain the information by local operation in a
deterministic manner. In this way, we complete the procedure of
secret extraction, and the above procedures from step 1 to step 4
constitute a complete process of secure QDC.

step 5. Repeat the above steps until all the secret information have
been transmitted.

\section{Discussion}
Now, we turn to the case if they do not choose to cooperate with
each other. Without the cooperation of Charlie, Bob still knows
which type Alice's operation belongs to, $(\textit{I},\sigma^{z})$
or $(\sigma^{y},i\sigma^{y})$. But he cannot further discriminate
which one Alice's operation is. Without the cooperation of Bob,
Charlie knows nothing about Alice's operation. If Charlie lies to
Bob, Bob also has a probability of $1/2$  to get the correct
information, so the successful cheat probability of Charlie is
$1/2$. Conversely, Charlie only has a probability of $1/4$ to get
the correct information, so the successful probability of Bob is
$3/4$. This is the point that we addressed above, he who received
Alice's atom has a higher probability of successful cheat compared
with the party who have not. If there is an eavesdropper, he may be
one of Bob and Charlie, or another one besides the three parties.
The eavesdropper has been able to entangle an ancilla with the GHZ
state, and at some later time she can measure the ancilla to gain
information about the measurement results of the users. However,
Hillery \textit{et al}. \cite{2} show that if this entanglement does
not introduce any errors into the procedure, then the state of the
system is a product of the GHZ triplet and the ancilla, which means
the eavesdropper could gain nothing about the measurements on the
triplet from observing his/her ancilla.

We also note the scheme can generalize to multi-users case providing
Alice possesses a multipartite entangled state. Suppose she has a
(\textit{N}+1)-qubit entangled state, qubits
$2,3,\cdot\cdot\cdot(\textit{N}+1)$ are to \textit{N} users,
respectively. After she confirms that each of the users have
received a qubit, she then operates one of the four local
measurements on qubit 1. After that, the two cbits information was
encoded into the (\textit{N}+1)-qubit entangled state. Later, she
sends her qubit to one of the rest \textit{N} users. Again, he who
received the qubit will have a higher probability of successful
cheat compared with the rest (\textit{N}-1) users. Only with the
cooperation of all the rest users, one can obtain Alice's
information. In this way, we set up a secure network for QDC via
QSS.

Next, we will give a brief analysis of the experimental feasibility
of our scheme. During the process, our scheme only involves
atom-field interaction with a large-detuned cavity and does not
require the transfer of quantum information between the atoms and
cavity. In addition, with the help of a strong classical driving
field the photon-number dependent parts in the evolution operator
are canceled. Thus the scheme is insensitive to both the thermal
field and the cavity decay. So, the requirement on the quality
factor of the cavities is greatly loosened. Meanwhile it is noted
that the atomic state evolution is independent of the cavity field
state, thus based on cavity QED techniques presently \cite{4,5} it
might be realizable. In our scheme, the two atoms must be
simultaneously interaction with the cavity. But in real case, we
can't achieve simultaneousness in perfect precise. Calculation on
the error suggests that it only slightly affects the fidelity of the
reconstruct state \cite{4}.

\section{Summary}
In summary, we have investigated a secure scheme for QDC with GHZ
type entangled state Via QSS in cavity QED. If and only if when they
cooperate with each other, they can read the original information.
Any attempt to get complete information of the state without the
cooperation of the third party cannot be succeed in a deterministic
way. In the scheme we provide a way to achieve all operations of
dense coding, from generation of the entangled state to various
measurements, by using cavity QED techniques. we The distinct
advantage of the scheme is that during the passage of the atoms
through the cavity field, a strong classical field is accompanied,
thus the scheme is insensitive to both the cavity decay and the
thermal field. Furthermore, our scheme only employs single-qubit
computational basis measurements, thus it may offer a simple and
easy way of demonstrating secure QDC experimentally in cavity QED
via QSS.

\begin{acknowledgments}
This work is supported by Anhui Provincial Natural Science
Foundation under Grant No: 03042401, the Key Program of the
Education Department of Anhui Province under Grant No: 2002kj029zd
and 2004kj005zd and the Talent Foundation of Anhui University.
\end{acknowledgments}

\end{document}